\documentclass[
reprint,
superscriptaddress,
showpacs,
showkeys,
 amsmath,amssymb,
 aps,
 prl,
]{revtex4-2}

\usepackage[normalem]{ulem}

\usepackage{xcolor}
\usepackage{upgreek}
\usepackage[T1]{fontenc}
\usepackage{graphicx}
\usepackage{dcolumn}
\usepackage{bm}
\usepackage{hyperref}
\usepackage{ulem}
\usepackage{subfigure}
\hypersetup{
    colorlinks=true,
    linkcolor=magenta,
    filecolor=red,      
    urlcolor=blue,
    pdftitle={Final \textsc{Conus} results on coherent elastic neutrino nucleus scattering at the Brokdorf reactor},
}
\usepackage[all]{hypcap}
\usepackage[inter-unit-product =\cdot]{siunitx}
\usepackage{gensymb}

\begin{document}

\title{Final \textsc{Conus} Results on Coherent Elastic Neutrino-Nucleus Scattering at the Brokdorf Reactor}

\makeatletter
\def\@fnsymbol#1{\ensuremath{\ifcase#1 \or \dagger \or \dagger\dagger \or \ddagger \or \ddagger\ddagger \or * \or \mathsection \or \mathsection\mathsection \or ** \or \parallel \or \mathparagraph \else\@ctrerr\fi}}
\makeatother

\newcommand{\MPIK}{\affiliation{Max-Planck-Institut f\"ur Kernphysik, Saupfercheckweg 1, 69117 Heidelberg, Germany}}
\newcommand{\PE}{\affiliation{PreussenElektra GmbH, Kernkraftwerk Brokdorf GmbH \& Co.oHG, Osterende, 25576 Brokdorf, Germany}}

\author{N.~Ackermann}\MPIK
\author{H.~Bonet}\MPIK
\author{A.~Bonhomme}\MPIK
\author{C.~Buck}\MPIK
\author{K.~F\"ulber}\PE
\author{J.~Hakenm\"uller}\MPIK
\author{J.~Hempfling}\MPIK
\author{J.~Henrichs}\MPIK
\author{G.~Heusser}\MPIK
\author{M.~Lindner}\MPIK
\author{W.~Maneschg}\MPIK
\author{T.~Rink}\MPIK
\author{E.~S\'{a}nchez Garc\'{i}a}\MPIK
\author{J.~Stauber}\MPIK
\author{H.~Strecker}\MPIK
\author{R.~Wink}\PE

\collaboration{CONUS Collaboration}
\email{conus.eb@mpi-hd.mpg.de}

\date{\today}

\begin{abstract}
The \textsc{Conus} experiment studies coherent elastic neutrino-nucleus scattering in four 1\,kg germanium spectrometers. Low ionization energy thresholds of 210\,eV were achieved. The detectors were operated inside an optimized shield at the Brokdorf nuclear power plant which provided a reactor antineutrino flux of up to 2.3$\cdot$10$^{13}$\,cm$^{-2}$s$^{-1}$. In the final phase of data collection at this site, the constraints on the neutrino interaction rate were improved by an order of magnitude as compared to the previous \textsc{Conus} analysis. The new limit of less than 0.34 signal events\,kg$^{-1}$\,d$^{-1}$ is within a factor 2 of the rate predicted by the standard model. This constraint is discussed in the context of conflicting measurements and results from another reactor neutrino experiment using similar technology.
\end{abstract}

\keywords{Nucleus-neutrino interactions, Semiconductor detectors, Nuclear reactors}

\maketitle

Coherent elastic neutrino-nucleus scattering (CE$\nu$NS) is a standard model (SM) interaction already predicted in 1974~\cite{Freedman:1973yd}. Its characteristic feature is a coherently enhanced cross section that is orders of magnitudes larger than inverse $\beta$ decay (IBD) and neutrino-electron scattering. This allows for a more compact design of neutrino detectors that offer more flexibility and mobility. The cross section of the CE$\nu$NS reaction increases with the number of neutrons squared favoring higher mass nuclei to get higher event rates. On the other hand, the reaction's imprint in a detector, i.e.~the nuclear recoil (nr) energy in the keV$_{nr}$ range, scales with the inverse of the mass of the target nucleus. Because of dissipation processes also known as quenching effects, the ionization energy observed in the detector, indicated by the unit eV$_{ee}$ (electron equivalent energy), is even smaller than the deposited recoil energy. Thus, in terms of event detection, the use of a medium atomic number target such as germanium (Ge) offers a good compromise between high statistics and experimental constraints.

CE$\nu$NS studies allow one to probe several SM and beyond the SM (BSM) physics processes. The first detection was achieved by the COHERENT Collaboration at a spallation neutron source, where neutrinos and antineutrinos up to 50\,MeV energy are produced in pion decays at rest~\cite{Coherent:2017, COHERENT:2020iec}. Another promising neutrino source considered to investigate CE$\nu$NS are nuclear reactors, which produce lower energetic antineutrinos, i.e.~$E_{\nu}\lesssim10$\,MeV. In this low-energy regime, the sensitivity to specific BSM channels is very high~\cite{CONUS:2021dwh, CONUS:2022qbb}. Compared to a typical IBD detector, a setup aiming to measure CE$\nu$NS is smaller and more adopted to the safety constraints in the inner zone of a nuclear power plant. Both facilitate installation at the site and make them promising devices for nuclear safeguard or reactor monitoring application in the future. Currently, there is a strong effort to detect CE$\nu$NS at reactors worldwide~\cite{CONNIE:2021ggh, nGeN:2022uje, NUCLEUS:2019igx, Colaresi:2022obx, Ricochet:2022pzj, Kerman:2016jqp} with the most stringent constraints provided by the \textsc{Conus} experiment~\cite{CONUS:2020skt}.

\begin{figure}
	\includegraphics[width=0.45\textwidth]{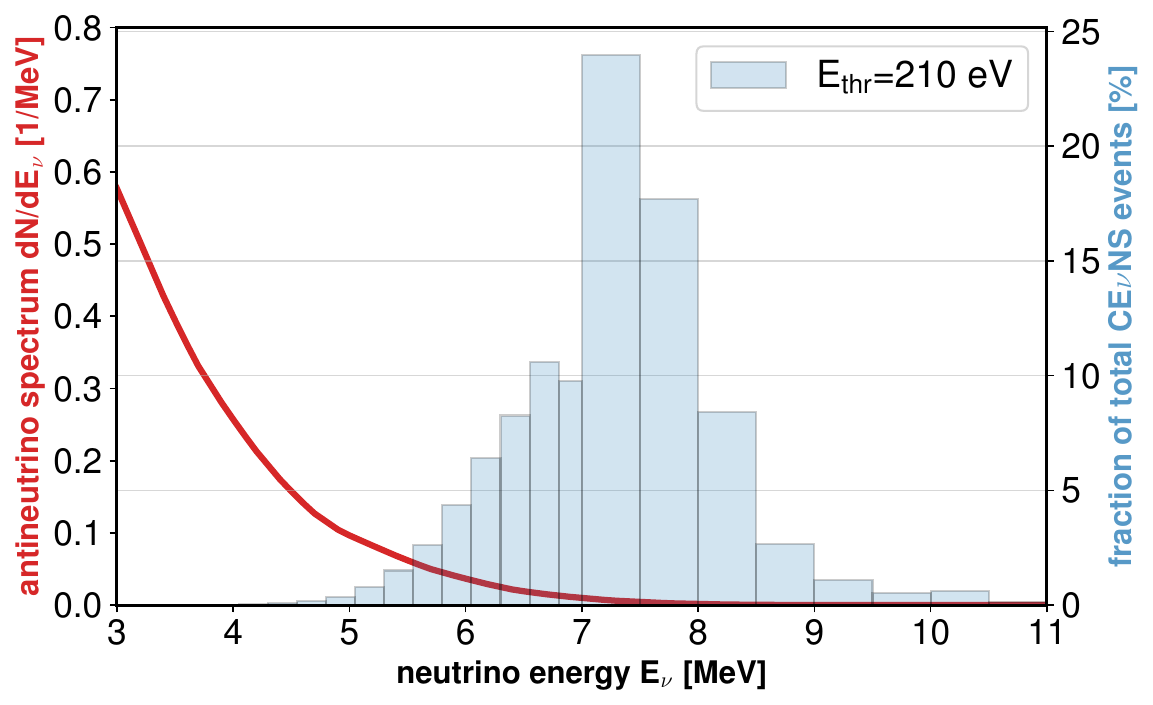}
	\caption{Reactor antineutrino spectrum (red) and the expected contribution to the number of CE$\nu$NS events above the analysis energy threshold (blue). Varying binning indicates the different measurements of the reactor's antineutrino spectrum that were combined for the \textsc{Conus} signal prediction. The applied spectrum extends up to $E_{\nu}$=11\,MeV.} 
	\label{fig:1}
\end{figure}

To be able to detect the tiny nuclear recoils in the CE$\nu$NS interaction, a technology assuring sub-keV$_{ee}$ energy thresholds and ultralow noise levels is required. For \textsc{Conus}, four p-type point-contact high-purity germanium (HPGe) detectors called C1--C4~\cite{Bonet:2020ntx} were utilized to fulfil these requirements. The lateral and top surface of the cylindrical HPGe diodes consist of a lithium-diffused n+ layer. On the bottom side, there is a large passivation layer and the p+ point contact used as readout electrode. The total active mass of the four detectors equipped with electrical cryocoolers is ($3.73\pm0.02$)\,kg~\cite{CONUS:2020skt}. However, one of the detectors (C3) had to be excluded from the analysis of this publication due to its high and unstable noise level. The detectors were placed in a compact shield with a total mass of 11\,tons. Five layers of lead, each with a thickness of 5\,cm efficiently suppressed the external $\gamma$ radiation. Plates of polyethylene, partly loaded with boron, were used to reduce the neutron flux. The thermal neutron flux at our experimental site is highly correlated with the reactor power. However, it was demonstrated that associated correlated background events in the region of interest (ROI) remaining after neutrons crossing all shield layers are negligible for all \textsc{Conus} analyses~\cite{Hakenmuller:2019ecb}. The shield also incorporated an active muon-veto system, consisting of plates of plastic scintillator and photomultiplier tubes. Finally, the detector chamber inside the shield was continuously flushed using breathing air bottles to keep the airborne radon concentration around the Ge detectors at a negligible level.

The \textsc{Conus} setup was positioned at a distance of 17.1\,m from the reactor core center at the nuclear power plant in Brokdorf, Germany (KBR)~\cite{Hakenmuller:2019ecb}. The maximum thermal power of this reactor which stopped operation by the end of 2021 was 3.9\,GW. The time-dependent fission fractions of the most relevant nuclides for antineutrino production, $^{235}$U, $^{238}$U, $^{239}$Pu and $^{241}$Pu, were provided by PreussenElektra GmbH, the company operating the plant. The averaged values are (49.1, 7.4, 36.1, 7.4)\%, respectively. For the prediction of the antineutrino spectrum, a data-based approach as proposed in~\cite{DayaBay:2021dqj} has been chosen. The recent Daya Bay measurement above 8\,MeV~\cite{DayaBay:2022eyy} was incorporated for the spectrum's high-energy part, while updated summation results of \cite{Estienne:2019ujo} were used to account for the antineutrino emission below the IBD threshold of $\sim$1.8~MeV. The predicted antineutrino spectrum and the expected contributions to the measured signal rate in the \textsc{Conus} detectors are shown in Fig.~\ref{fig:1}.
 
Whereas the digital multichannel analyzer Lynx~\cite{Bonet:2020ntx} was used in the first phase of the data collection in \textsc{Conus} (Runs 1--4), a data acquisition system (DAQ) based on CAEN electronic modules has been used for the \textsc{Conus} Run-5 data discussed in this publication. This dataset includes a reactor on period in 2021 and a reactor off period in 2022 allowing for a precise background determination. The new DAQ system has the advantage of recording the charge collection pulses offering the possibility for background rejection by pulse-shape discrimination (PSD) on an event-by-event basis. The signals of the HPGe detectors are digitized by a CAEN V1782 multichannel analyzer which is operated by the CAEN CoMPASS software. Detector signals are triggered by a combination of a slow and a fast triangular discriminator that allows one to obtain a better trigger efficiency than the standard RC-CR2 algorithm. More information can be found in~\cite{bonet2023pulse}. The logic signals of the muon-veto system and the transistor reset preamplifier (TRP) of the HPGe detectors are recorded by a CAEN V1725 module. Both modules are synchronized in time allowing for an off-line application of anticoincidence time cuts between Ge data and logic signals.

With the new DAQ, a PSD cut was developed for the \textsc{Conus} experiment \cite{bonet2023pulse}. With its application, background rejection down to the energy threshold in the sub-keV$_{ee}$ regime became possible. Based on the shape of the detector pulses, in particular their rise time, background events at the Ge diode's surface can be distinguished from events generated in the bulk volume. For the Run-5 CE$\nu$NS analysis, a conservative PSD cut with a signal acceptance of $97\%$ was applied, which leads to a background reduction in the ROI ranging from 0.21 to 1.00\,keV$_{ee}$ by 5\%--10\%. Therein, a rejection of surface events of about $57\%$ on average is achieved. An additional benefit of the applied PSD cut is its impact on the shape of the background spectrum. As a surface event contamination is expected to rise toward low energies in almost the same manner as a CE$\nu$NS or BSM signal, efficient reduction of this contribution proved to be beneficial for the outcome of this analysis.  

As compared to the previous analysis~\cite{CONUS:2020skt} two new background components were considered for the background description of Run-5. The first one takes into account the additional background contribution induced by a leakage test at the end of the reactor outage of Run-2 (June 2019). Since the reactor containment vessel was set to overpressure for this test, the vacuum inside the cryostats was replaced by pure gaseous argon as a safety measure~\cite{Bonet:2020ntx}. Subsequently, a continuous detector-dependent enhancement in the background below 10\,keV$_{ee}$ was observed of up to a factor of 2. No indication of a radioactive contamination was found. Thus, this component was incorporated as an analytic function parametrized with two free parameters added to the previous background model~\cite{Bonet:2021wjw}. In the resulting updated model, the normalization of the background (referred to as b parameter in~\cite{Bonet:2021wjw}) is limited to an order of magnitude of 10\% by the statistics collected prior to the leakage test. The extended reactor off period of Run-5 enabled us to obtain a high-statistics sample of the background. Comparing these data to the reactor on period revealed an additional second background contribution during reactor on times of about 1-2\,counts\,kg$^{-1}$\,d$^{-1}$ ($<6$\% of total background), caused by a change in veto efficiency due to the high-energetic reactor-correlated $\gamma$ radiation (predominantly $^{16}$N~\cite{Bonet:2021wjw}) detected by the muon-veto system. The contribution of the high-energetic $\gamma$-ray background itself was found to be negligible as estimated from simulations.

\begin{figure}
	\includegraphics[width=0.45\textwidth]{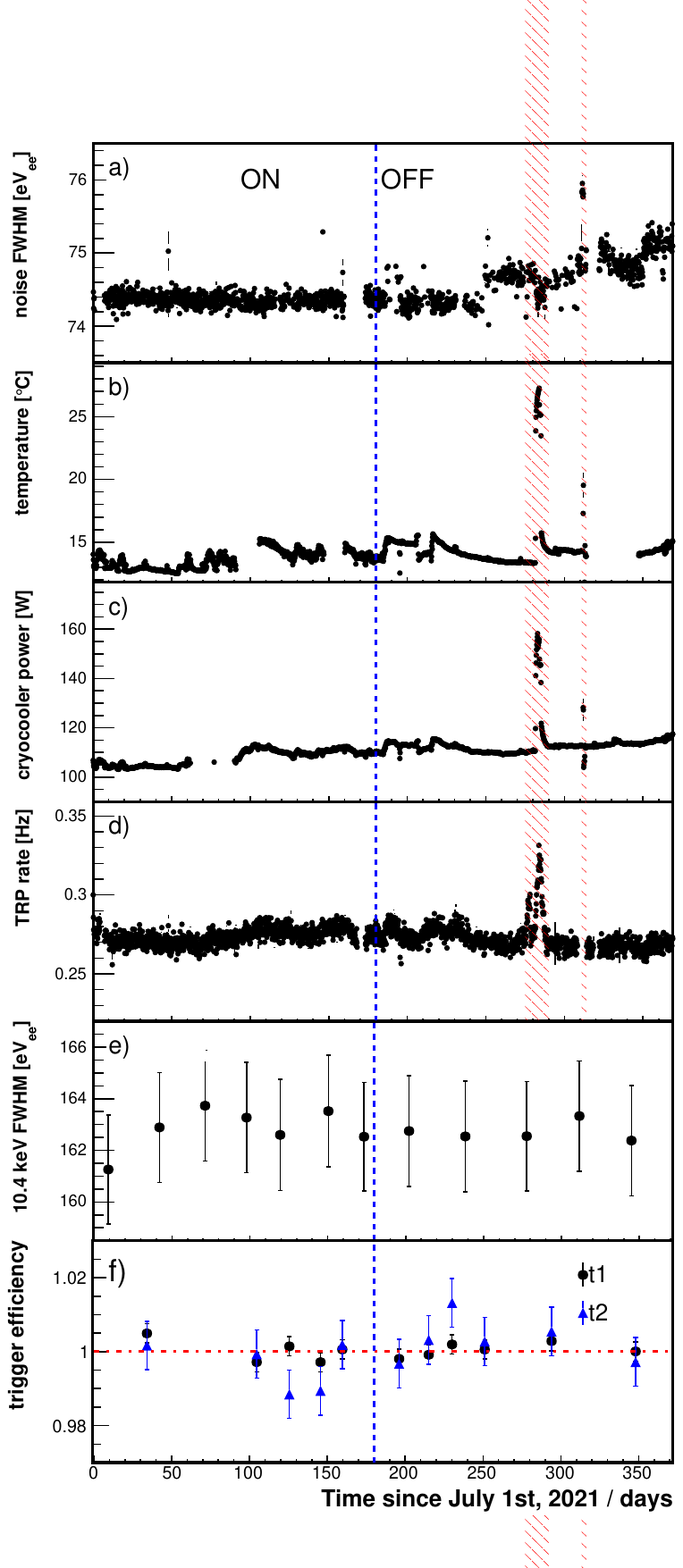}
	\caption{Evolution of parameters for the C2 detector during Run-5. (a) FWHM of noise peak, (b) air temperature close to the detector preamplifier, (c) cryocooler power consumption, (d) TRP trigger rate, (e) FWHM of K-shell EC peak of $^{71}$Ge, (f) relative variation of trigger efficiency curve parameters~\cite{bonet2023pulse}.}
	\label{fig:2}
\end{figure}

The evolution of the main environmental parameters during reactor on and off is shown in Fig.~\ref{fig:2} for C2. The peak full width at half maximum (FWHM) from noise events triggering the DAQ was found to be Gaussian in good approximation and its width was monitored over time. Time periods with variations of the FWHM over 1\,eV$_{ee}$ were removed (red-shadowed area in Fig.~\ref{fig:2}). These periods are mainly related to large temperature changes in the room when the air conditioning system failed. At the same time, the cryocooling system increases its power to compensate for these variations, producing more vibrations and thus more microphonic events. During normal operation the room temperature was stable with variations below 2\,$^{\circ}$C and an average mean value of ($14.4\pm0.7)^{\circ}$C. Additional quality cuts based of the time difference between single events and the reset rate of the preamplifier (TRP) are applied to remove noise and spurious events as in the previous analysis~\cite{Bonet:2020ntx}. The FWHM evolution of the peak at 10.37\,keV$_{ee}$ from K-shell electron capture (EC) in $^{71}$Ge is displayed in Fig.~\ref{fig:2} with variations below 2\,eV$_{ee}$. Finally, the development of the trigger efficiency curve parameters (described in~\cite{bonet2023pulse}) during Run-5 is depicted, showing differences of less than 2\%.  
 
These highly stable environmental conditions allow one to lower the energy threshold from about 300\,eV$_{ee}$ to 210\,eV$_{ee}$ compared to previous analyses. The energy threshold is defined independently for each detector based on two different criteria: First, that the trigger efficiency is over 20\%, and second the contribution of the noise peak to the ROI is below 1~count\,kg$^{-1}$\,d$^{-1}$\,keV$_{ee}^{-1}$. An energy threshold of 210\,eV$_{ee}$ is defined for the C1, C2 and C4 detectors following this approach. The upper cut defining the energy ROI was kept at 1\,keV$_{ee}$. Taking into account the active detector masses, an exposure of 426\,kg~d during reactor on and 272\,kg\,d during reactor off was obtained. The individual contributions from the three detectors are listed in Table~\ref{tab:1}.

\begin{table}
\caption{\label{tab:1} Reactor on and off exposure for the three detectors used in the Run-5 analysis after cuts, taking into account the active masses.}
\centering
\begin{tabular}{|c|cc|}\hline
Detector & Reactor on [kg d] & Reactor off [kg d] \\\hline
C1 & 141.5 & 40.2 \\
C2 & 145.5 & 130.3 \\
C4 & 139.0 & 101.6 \\\hline
Total & 426 & 272 \\\hline
\end{tabular}
\end{table} 
@article{Ackermann:2024kxo,
    author = "Ackermann, N. and others",
    title = "{Final CONUS results on coherent elastic neutrino nucleus scattering at the Brokdorf reactor}",
    eprint = "2401.07684",
    archivePrefix = "arXiv",
    primaryClass = "hep-ex",
    month = "1",
    year = "2024"
}
The dead time induced by the muon-veto system was estimated independently for reactor on and off periods using an average veto window of 450\,$\mu$s. The dead time was measured to be on average ($7.00\pm0.25$)\% (reactor on) and ($3.39\pm0.10$)\% (reactor off). The dead time increased when the reactor was operating due to the enhanced reactor-correlated $\gamma$-ray flux~\cite{Bonet:2021wjw}. The averaged dead time induced by the DAQ for the HPGe detectors is calculated to be ($1.07\pm0.10$)\%. 

\begin{figure}
	\includegraphics[width=0.45\textwidth]{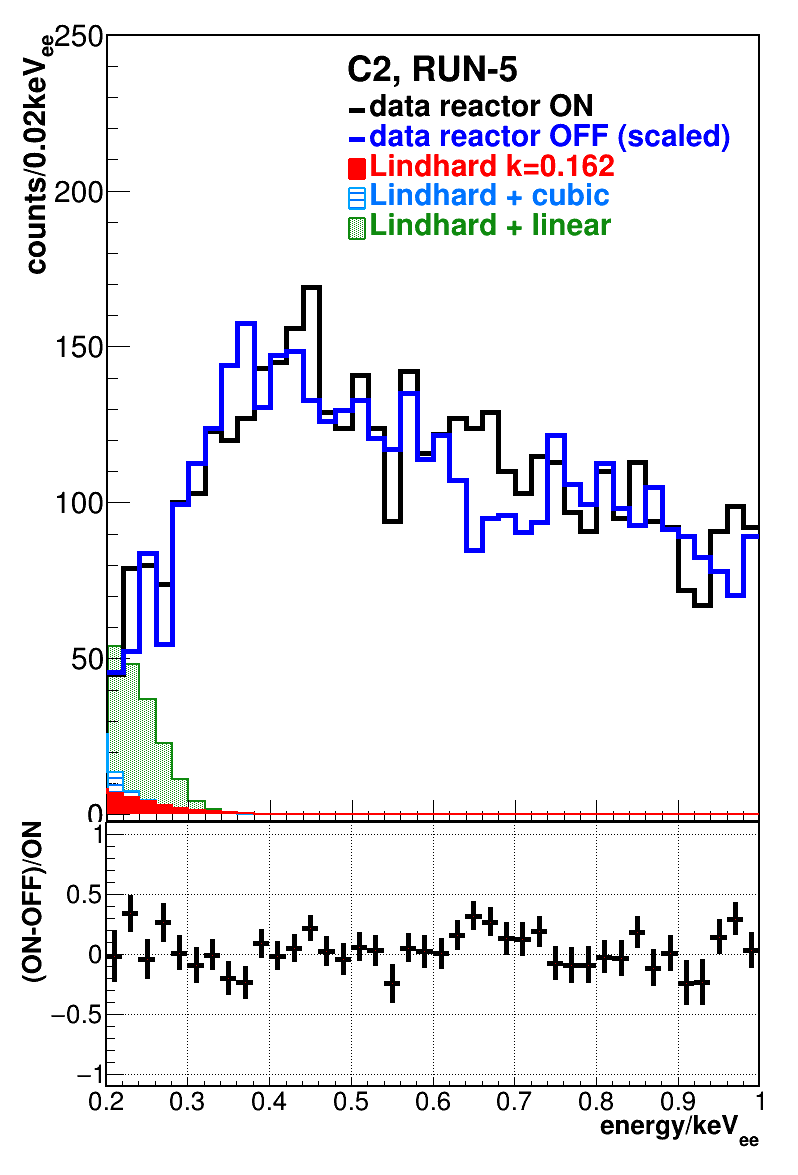}
	\caption{Reactor on vs reactor off data for the C2 detector taken in Run-5 of the \textsc{Conus} experiment  including the predicted CE$\nu$NS events of the quenching descriptions listed in Table~\ref{tab:3}.}
	\label{fig:3}
\end{figure}

Figure~\ref{fig:3} shows the measured reactor on and reactor off spectra for the C2 detector. As the collected data do not show any significant excess above reactor off background, an upper limit on the number of CE$\nu$NS events is determined. 

Thus, the statistical analysis is based on a profile likelihood ratio test~\cite{Cowan:2010js}, in which the individual detector datasets of reactor on and reactor off periods are fitted simultaneously. Underlying experimental uncertainties were considered with detector-specific (Gaussian) pull terms in the combined likelihood function. An upper limit on the total number of CE$\nu$NS events was deduced from a $\chi^{2}$-distributed (one-sided) test statistics. 

\begin{figure}
	\includegraphics[width=0.45\textwidth]{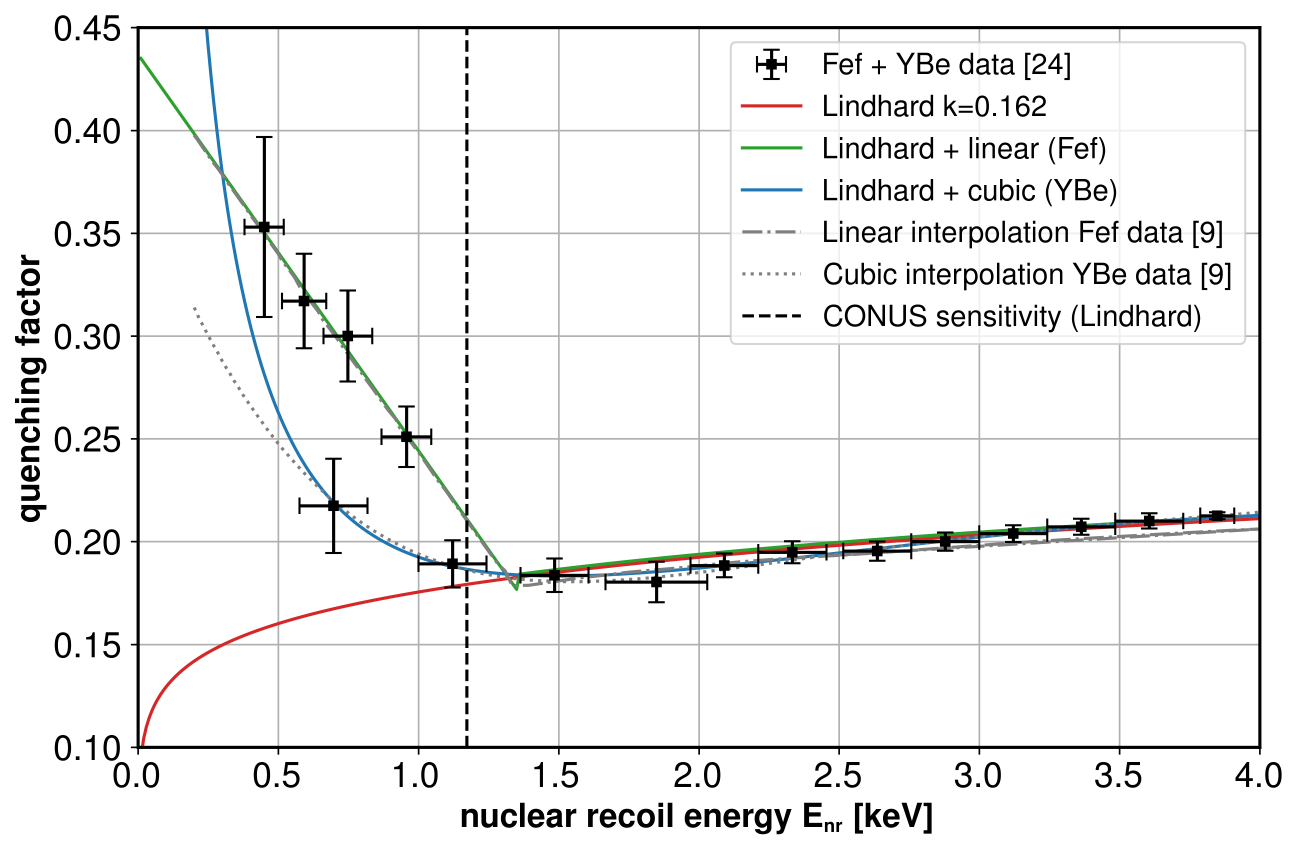}
	\caption{Quenching descriptions (colored lines) used in the analysis of this work. The low-energy rise has been measured in~\cite{Collar:2021fcl}, while the Lindhard model has been confirmed by~\cite{Bonhomme:2022lcz, Liphd}. For comparison, curves related to the individual quenching measurements -- iron-filtered (Fef) and photo-neutron (YBe) data -- given by \cite{Colaresi:2022obx} are shown in gray. The mean effective threshold of the \textsc{Conus} detectors is represented by a vertical dashed line, assuming the Lindhard model.}
	\label{fig:4}
\end{figure}

The most critical contributions to the systematic uncertainties originate from the trigger efficiency estimation and signal quenching in the HPGe detectors given by the ratio of ionization signal for nuclear recoils over the signal from electrons of the same energy. The quenching effect can be described by the Lindhard theory~\cite{lindhard1963range}. Good agreement between experimental data and the Lindhard prediction was found recently in~\cite{Bonhomme:2022lcz, Liphd}. In another new measurement~\cite{Collar:2021fcl}, contradicting results were obtained featuring an unexpected enhancement of the quenching factor in the sub-keV$_{ee}$ region and deviating significantly from the Lindhard model. In this scenario, the signal expected in \textsc{Conus} would be larger than in the case of the Lindhard prediction. The Migdal effect was proposed as a physical explanation for the enhancement. However, a recent theoretical study demonstrated that the Migdal contribution is negligible for the currently running CE$\nu$NS searches at nuclear reactors with Ge as target material~\cite{AtzoriCorona:2023ais}. Nevertheless, we also incorporated these quenching measurements from two different datasets named hereafter iron-filtered and photo-neutron data~\cite{Collar:2021fcl} in our analysis in order to test them independently with our Run-5 data. Following~\cite{Collar:2021fcl}, a linear fit has been used for the iron-filtered data below 1.35\,keV$_{nr}$, whereas a cubic-like fit was used for the photo-neutron data. In these fits, the errors on both axes were taken into account. An overview of the quenching descriptions considered in this analysis are depicted in Fig.~\ref{fig:4}. The systematic uncertainties related to energy calibration and the PSD parameters have small or negligible impact on the result.

As listed in Table~\ref{tab:3}, the predicted combined number of CE$\nu$NS events in the three detectors of the Run-5 dataset is ($91^{+11}_{-9}$) assuming a quenching following the Lindhard theory with a quenching parameter of $k=(0.162\pm0.004)$ as determined in~\cite{Bonhomme:2022lcz}. From the combined likelihood fit we obtain an upper limit of 143 antineutrino events at 90\% confidence level (CL), which corresponds to $< 0.34$ signal events\,kg$^{-1}$\,d$^{-1}$ and is less than a factor 2 above the SM expectation. This corresponds to a sensitivity improvement of an order of magnitude as compared to the previous \textsc{Conus} analysis~\cite{CONUS:2020skt} with Run-1 and Run-2 data and represents the current world-best limit for reactor experiments. Our new result strongly profits from the lower energy threshold of 210\,eV$_{ee}$ and the higher reactor off statistics.   

\begin{table}
\caption{\label{tab:3} Predicted signal events for the Run-5 exposure of 426\,kg\,d, experimental constraints for different quenching descriptions extracted from a one-sided hypothesis test and p values for the signal hypotheses.}
\centering
\begin{tabular}{|l|c|c|c|}\hline
Quenching description & Prediction & Run-5 limit & p value\\
 & & (90\% CL) & \\ \hline
Lindhard ($k=0.162$~\cite{Bonhomme:2022lcz}) & $91^{+11}_{-9}$ & 143 & 0.46\\
Linear low E excess~\cite{Collar:2021fcl} & $645^{+59}_{-90}$ & 99 & $<10${-5}\\
Cubic low E excess~\cite{Collar:2021fcl} & $115^{+13}_{-11}$ & 122 & 0.21\\\hline
\end{tabular}
\end{table} 

In Table~\ref{tab:3}, additional signal predictions and constraints derived from the \textsc{Conus} data are listed assuming an enhanced quenching factor toward lower energies. From this table, it can be seen that descriptions following the quenching data deviating from the Lindhard theory, in particular the ones determined with the iron-filtered neutron beam preferred in~\cite{Colaresi:2022obx}, lead to predictions for the antineutrino interaction rate well above the \textsc{Conus} limit. This indicates either a bias in part of the quenching measurements of~\cite{Collar:2021fcl}, an underestimated reactor-correlated background in~\cite{Colaresi:2022obx}, or BSM physics beyond the CE$\nu$NS rate estimated in~\cite{Colaresi:2022obx}. The quenching data extracted from the use of a $^{88}$Y/Be photo-neutron source in~\cite{Collar:2021fcl} represented by the cubic fit in Fig.~\ref{fig:4} cannot be excluded yet, despite some tension with the measurements in~\cite{Bonhomme:2022lcz}. There is an additional data point at $E_{nr}=254$\,eV~\cite{Collar:2021fcl} obtained from nuclear recoils after $\gamma$ emission following a thermal neutron capture. A recent measurement confirmed here a quenching factor higher than predicted by the Lindhard theory of $(25\pm2)$\,\%~\cite{Kavner:2024xxd}. However, this data point is well below the threshold of the CONUS detectors and therefore not relevant for our analysis. Moreover, this data point is not following the trend of the linear extrapolation of the iron-filtered data and is well below the green line of Fig.~\ref{fig:4}. Independent of the correct quenching factor, from a comparison of the experimental parameters in \textsc{Conus} and in the experiment at the Dresden-II reactor such as background level, energy resolution, threshold and statistics, a signal should have been seen in the \textsc{Conus} data of this work in case the excess reported in~\cite{Colaresi:2022obx} originates from CE$\nu$NS and/or BSM physics (see Appendix). 

Finally, following the Lindhard description and the SM CE$\nu$NS prediction, quenching parameters $k>0.21$ can be excluded at 90\% CL,from the absence of a neutrino signal in \textsc{Conus} Run-5 data.

In this work, the analysis of the final high-statistics \textsc{Conus} dataset collected at the nuclear power plant in Brokdorf was presented. The criticisms raised in~\cite{Collar:2024syc} on this analysis is discussed in the Appendix. Significantly lower energy thresholds combined with very stable environmental conditions allowed us to improve the upper limit on CE$\nu$NS of reactor antineutrinos by 1 order of magnitude, setting this limit only a factor 2 above the SM prediction, a Lindhard description of quenching assumed. The quality of this dataset is expected to further improve the bounds of our BSM studies. Investigations of neutrino electromagnetic properties and new nonstandard neutrino interactions are to appear soon. 

The \textsc{Conus} setup moved recently to its new location in Leibstadt, Switzerland (KKL). There, the detectors are operated next to a boiling water reactor of 3.6\,GW thermal power at a distance of about 20.7\,m from the center of the reactor core. Several improvements were realized at this new site. The HPGe detectors were refurbished to further improve the energy resolution and the trigger efficiency at low energy. This will allow for an even lower energy threshold. Furthermore, the muon-veto system was improved for a higher tagging efficiency. The new \textsc{Conus}+ experiment~\cite{CONUS:2024lnu} has just started data collection in Leibstadt and will continue to study the capabilities of this technology for fundamental research. 


We thank all the technical and administrative staff who helped in building the experiment, in particular the MPIK workshops and Mirion Technologies (Canberra) in Lingolsheim. We express our gratitude to the PreussenElektra GmbH for great support and for hosting the \textsc{Conus} experiment. 

\bibliography{references}{}

\clearpage

\section*{Appendix}

\subsection{Background model}
On top of the background model of the Run-1/2 data two extra components had to be added for the Run-5 data at KBR. One additional background source was introduced by the breaking of the vacuum of the cryostats to avoid their implosion during a regular leakage test of the reactor safety vessel on $5/6$~July~2019. Another background component was present when the reactor was on, likely due to different efficiencies of the muon veto in reactor on and off periods. 

\begin{figure}
	\includegraphics[width=0.45\textwidth]{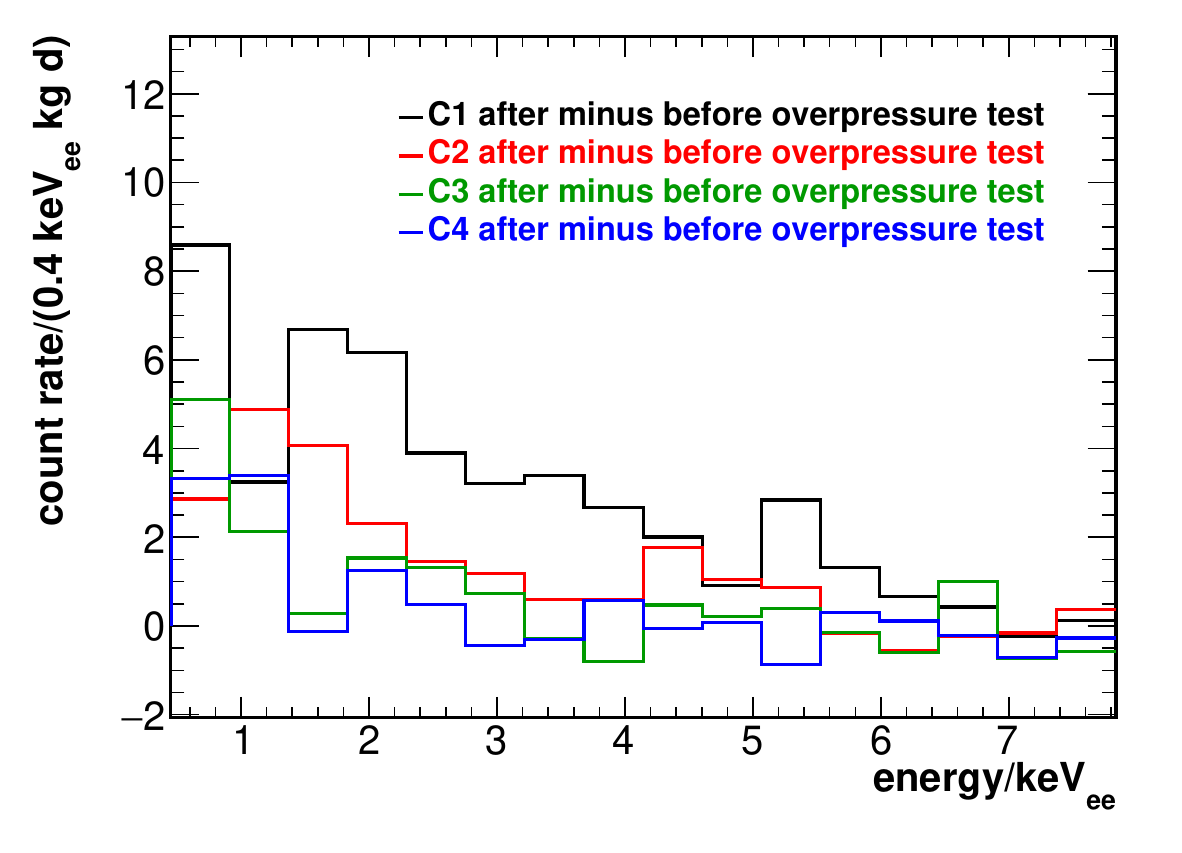}
	\caption{Impact of the leakage test on the spectra for C1 to C4.}
	\label{fig:appendix_leakagetest}
\end{figure}

The background component related to the leakage test is restricted to energies below 10\,keV$_{ee}$ and the impact was different for the four detectors with the largest increase of the count rate observed in C1 (see figure~\ref{fig:appendix_leakagetest}). In particular, the background rates above 10\,keV$_{ee}$ are fully compatible to the data collected before. 

Large efforts were taken to uncover the origin of this background component. From hardware point of view, several aspects were investigated:
\begin{itemize}
\item A change of the energy and trigger filter settings of the original Lynx DAQ did not have any impact on the background contribution.
\item An overall visual inspection of the pulse shapes with the new CAEN DAQ did not reveal any particularities.
\item A change in diode temperature by a few degrees, which is possible with the CP5 electric cryocoolers, did also not lead to a change in the background rate.
\item \textsc{Conus} detector irradiations with a $^{228}$Th calibration source revealed that data taken before and after the leakage test are in agreement. This indicates that the rate increase is related to an additional background contribution and not to an incorrect processing of the signals.
\end{itemize}

From a potential detector contamination point of view, different hypotheses were considered and rejected:

\begin{itemize}
\item The dismounting of the \textsc{Conus} detectors at KBR in 2022 and their upgrade for \textsc{Conus}+ at KKL offered the possibility to perform $\gamma$-ray screening measurements and wipe tests of all inner and outer copper parts around the Ge crystals. For the inner parts (crystal holder etc.) a surface contamination in U, Th, $^{40}$K, $^{137}$Cs and $^{60}$Co above $\sim30\,\mu$Bq\,cm$^{-2}$, $\sim30\,\mu$Bq\,cm$^{-2}$, $\sim80\,\mu$Bq\,cm$^{-2}$, $\sim10\,\mu$Bq\,cm$^{-2}$ and $\sim7\,\mu$Bq\,cm$^{-2}$ respectively was excluded. For the outer parts (end caps) a surface contamination below 10\,mBq\,cm$^{-2}$ was estimated for all isotopes except $^{210}$Pb with a sensitivity of 100\,mBq\,cm$^{-2}$.
\item Residual gas analysis of C2 during pumping to improve the vacuum: no unexpected isotopes were found and from the expected isotopes none were observed in noticeable quantities.
\item MC simulation of different isotopes with contributions at low energy: MC simulations of $^{210}$Pb, $^{60}$Co and $^{137}$Cs in the vicinity of the diode inside the cryostat were compared to the \textsc{Conus} detector data. While it is possible to reproduce the excess at low energies, all of these contributions would lead to a significant increase of the background at higher energies, for example in the 46.5\,keV line in the case of $^{210}$Pb. This was not observed in the data.
\item The stability of the count rate of the new background component was investigated. In the three years of operation after the leakage test, no reduction or increase in the count rate has been observed, which excludes short-lived radio-isotopes with half-lives of less than ten years (extrapolated from C1 with the largest contribution). This excludes the presence of $^{60}$Co, but not of $^{137}$Cs, both present as surface contamination at the experimental site. However, our MC simulations reveal, that also $^{137}$Cs placed in different detector positions cannot describe the new spectral component.
\end{itemize}

So, the origin of the new background component remains unknown. For the Run-5 analysis, this first additional background component was modeled using a function with two parameters added to the original background model. It was shown that choosing a different function leads to a compatible background description within the fit uncertainties. No significant impact on the determination of the CE$\nu$NS limit is expected.

Regarding the second additional background component, the long and stable reactor off period without any contributions of airborne radon decays to the background enabled a more detailed comparison of the reactor on to off data than it was possible before. This comparison revealed a small additional background contribution during Run-5 reactor on time. It amounts to 1-2 counts\,kg$^{-1}$\,d$^{-1}$ in the region of interested and the origin can be traced back to the muon-induced background. Our previous analysis was not sensitive to such
a background component due to statistical limitation. However, in Run-5, the muon veto window was prolonged and applied in a different way due to the DAQ upgrade.   

The rate difference between reactor on and off periods in the high energy range up to 500\,keV$_{ee}$ collected with C3 was compared with MC simulations of the muon-induced background as well as with high energetic $\gamma$-radiation of several MeV propagated through the shield. It was found that the shape of the muon-induced spectrum is a better fit to the data. Scaling the component to the high energy background furthermore reveals that only the muon-induced component contributes enough at low energies to explain the additional counts there. Thus, we decided to model this component with the muon-induced MC spectrum. We assume that the origin is related to differences in veto performance due to the presence of high energetic $^{16}$N $\gamma$-radiation created by the reactor operations. Additionally, we examined the impact on our result by excluding the component from the likelihood fit. This impact was found to be less than 10\%, since the shape of this background is very different from the expected CE$\nu$NS signal for which no contribution above 500\,eV$_{ee}$ is expected.  

The final major adaption in the background model stems from the application of the PSD to the data. In general, as described in \cite{bonet2023pulse}, the reduction of the background by the removal of the slow pulses is reproduced by the MC simulations (by removing the addition of the slow pulses in the post-processing and applying the respective efficiency). The impact of the PSD cut on the additional background induced by the leakage test is however not known. The PSD cut was applied to the background model including the leakage test fit function by scaling the full model with the energy-dependent background suppression observed in the data. This factor was obtained by comparing the spectra with and without the application of the PSD. The respective uncertainty is incorporated in a pull term of the likelihood function. 

\begin{figure}
	\includegraphics[width=0.45\textwidth]{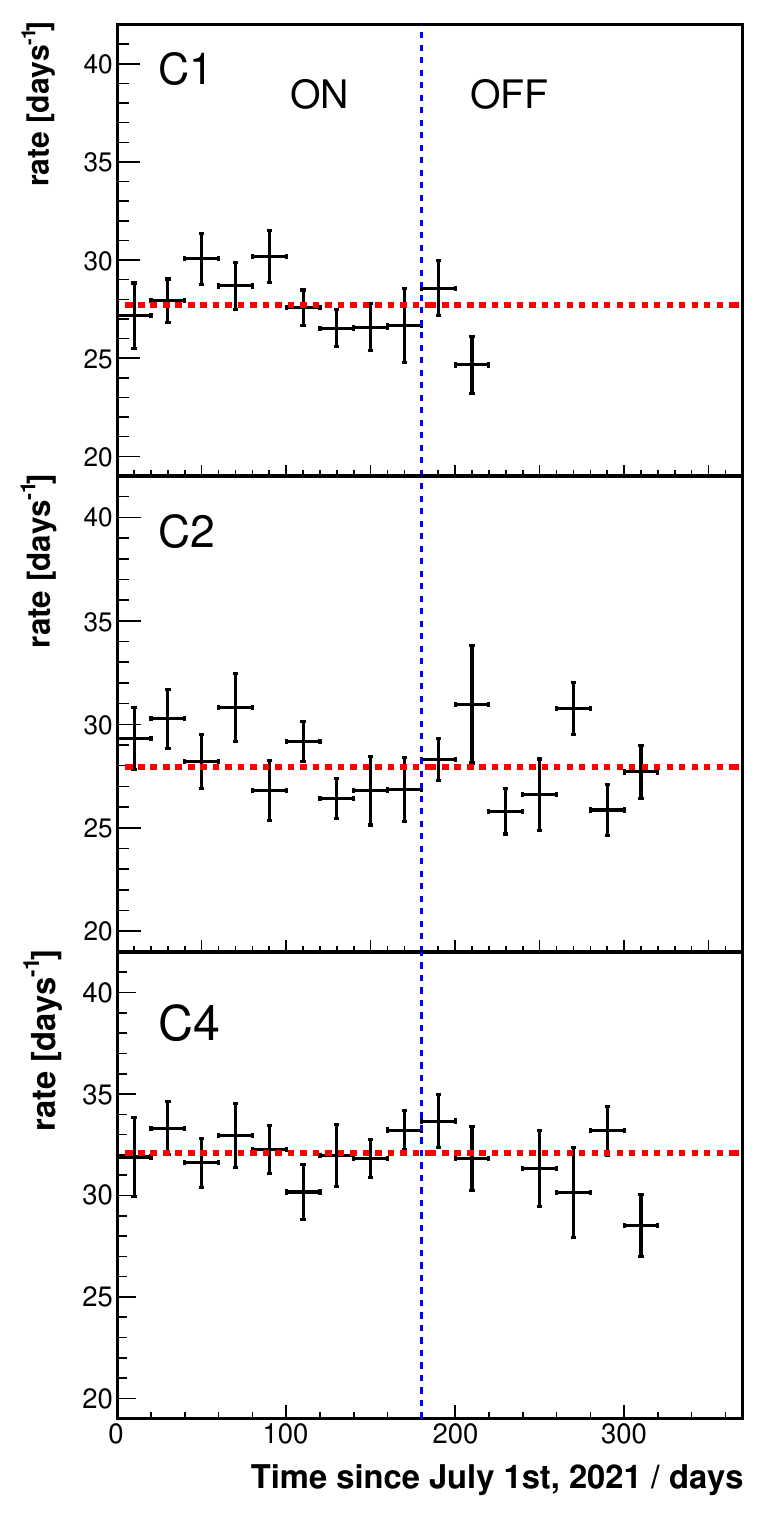}
	\caption{Rate evolution without trigger efficiency correction during Run-5 in the [0.21-1]~keV$_{ee}$ energy region. The rate values are averaged over 20 days. The average value is shown for each detector with the dashed line (red). Small differences are observed between reactor on and off data.}
	\label{fig:bkg}
\end{figure}

Finally, the validity of the updated Run-5 background model was confirmed by the flat residuals of the likelihood fit. The overall stability of the background and event rate in the energy region of interest without correction of the trigger efficiencies is demonstrated in figure~\ref{fig:bkg}.

\begin{figure}
	\includegraphics[width=0.5\textwidth]{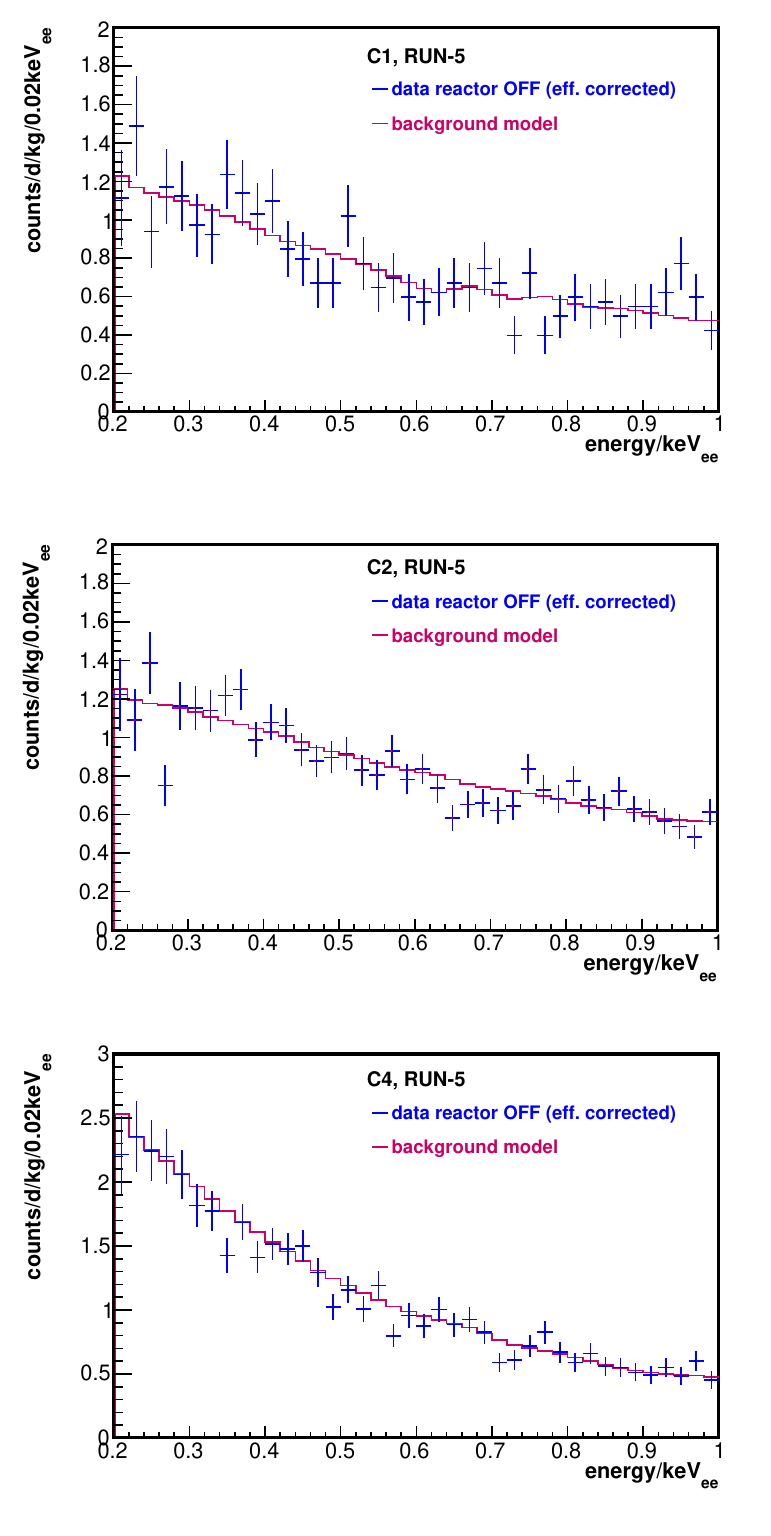}
	\caption{Comparison of the background model and reactor off data for the 3 CONUS detectors. The error bars for the data points are statistical only.}
	\label{fig:bkgvsdataROI}
\end{figure}

In figure~\ref{fig:bkgvsdataROI} our background model is compared to the reactor off spectrum in the energy region of interest. The spectrum was corrected for the trigger efficiency. This figures demonstrates the compatibility of our reactor off data with the physical background model. In particular, it shows that there is no significant contribution of microphonic noise events to the spectrum. The differences in the spectral shapes for the individual detectors can be explained by the different contributions of e.g.~$^{210}$Pb within the cryostat or the additional component after the leakage test to the overall spectrum~\cite{Bonet:2021wjw}.

In previous publications, we reported a significant contribution of noise events at and around the former analysis threshold of about 300\,eV$_{ee}$, partly induced by vibrations of the cryocooler systems~\cite{Bonet:2020ntx, CONUS:2021dwh}. Therefore, the setup was rather sensitive to temperature and cryocooler power variations during Run-1 and Run-2. For the Run-5 dataset discussed in this publication, the noise level was well below the background rate described in our background model and the noise variations can be neglected as compared to the expected signal rates. This major improvement is related to our new DAQ system used in Run-5 and the lower, more stable room temperature. To make sure that higher temperatures in the CONUS room and associate higher values of the cryocooler power do not induce a significant rate increase, we analysed data periods removed from the dataset when the air conditioning system was switched off. In such periods, the relative increase in cryocooler power was an order of magnitude higher as compared to the reactor ON/OFF differences. No significant change in the event rate stability was observed even under these extreme conditions. Therefore, we can exclude that noise variations from fluctuations of the environmental parameters significantly enhance the event rate in the reactor off data, encumbering the observation of a signal excess as speculated in~\cite{Collar:2024syc}.

\subsection{Quenching models}
For signal quenching in the \textsc{Conus} HPGe detectors, the conventionally used model of Lindhard is applied~\cite{lindhard1963range}. It describes the energy dependent ratio between nuclear recoil energy $E_{\mathrm{nr}}$ and detectable ionization energy $E_{\mathrm{ee}}$ as given by the following form.
    \begin{align}
        f(E_{\mathrm{nr}}; k) = \frac{k\, g(\epsilon)}{1+ k\, g(\epsilon)}\, 
    \end{align}
    with the nuclear recoil energy $E_{\mathrm{nr}}$ in keV, the 'quenching' parameter $k$, $\epsilon(E_{\mathrm{nr}})=11.5\, Z^{-7/3} E_{\mathrm{nr}}$ and $g(\epsilon)=3\epsilon^{0.15} + 0.7 \epsilon^{0.6} + \epsilon$. Here, the quenching parameter $k$ indicates the conversion efficiency at $E_{\mathrm{nr}}\sim1$\,keV. 
    Furthermore, to describe the quenching measurement of~\cite{Collar:2021fcl}, we follow the instructions of the corresponding authors and fit their data with two separate functional forms modifying the Lindhard model at lower energies, i.e.\ a linear and a cubic function, respectively. Uncertainties on both x-axis and y-axis have been incorporated in the fitting procedure. In the following, we give the used functional expression together with their best-fit values:
    For the iron-filtered (Fef-) data we apply
    \begin{align}
        \text{QF}(E_{\mathrm{nr}}) = \begin{cases}
            m\cdot E_{\mathrm{nr}} + n\, , & \text{for } E_{\mathrm{nr}}\leq 1.35\,\text{keV}\, ,\\
            \quad f(E_{\mathrm{nr}})\, , & \text{for } E_{\mathrm{nr}}> 1.35\,\text{keV}\, ,
        \end{cases}
    \end{align}
    with the best-fit values $k=0.164$, $m=-0.193$, $n=0.437$, while the photo-neutron (YBe) data are described by  
    \begin{align}
        \text{QF}(E_{\mathrm{nr}}) = \begin{cases}
            (a E_{\mathrm{nr}}^{3} + b E_{\mathrm{nr}}^{2} + c E_{\mathrm{nr}} + d)/E_{\mathrm{nr}}, & \\ \text{for } E_{\mathrm{nr}}\leq 3.5\,\text{keV}\, ,\\
            f(E_{\mathrm{nr}})\, , & \\ \text{for } E_{\mathrm{nr}}> 3.5\,\text{keV}\, ,
        \end{cases}
    \end{align}
    with the parameters $k=0.164$, $a=-0.006$, $b=0.060$, $c=0.042$, $d=0.096$.
    
\subsection*{Likelihood fit and systematics treatment}

The likelihood function used to determine the upper limits on the CE$\nu$NS rate for the Run-5 data is 

\begin{equation}
    \begin{aligned}
        -2\log\mathcal{L} & = -2\log\mathcal{L}_{\mathrm{ON}} -2\log\mathcal{L}_{\mathrm{OFF}} \\
        & + \sum_{i} (\bf{\theta}_{i} - \bf{\bar{\theta}}_{i})^T \ \text{Cov}_{i}^{-1}\  (\bf{\theta}_{i} - \bf{\bar{\theta}}_{i})\\
        & + \sum_{i} \frac{(\theta_{i} - \bar{\theta}_{i})^2}{\sigma_{\theta_{i}}}\, ,
    \end{aligned}
\end{equation}

 including the likelihood functions for the reactor on and off periods, $\mathcal{L}_{\mathrm{ON}}$ and $\mathcal{L}_{\mathrm{OFF}}$ respectively and the pull terms for the systematic uncertainties. Here, the first term represents the pull terms of correlated parameters, namely the pulse shape discrimination parameters (background rejection and the signal acceptance) as well as the trigger efficiency. The second term incorporates external knowledge of individual parameters, like the active mass of the detectors, the reactor neutrino flux and the uncertainty on the energy scale calibration of the spectra. 

 Overall, the signal and background contributions depend on a multitude of fit parameters:

\begin{equation}
    \begin{aligned}
        S=S(s, \phi, m_{\mathrm{act}}, \mu_{\mathrm{eff}}, \sigma_{\mathrm{eff}},\mu_{\mathrm{PSD}}, \sigma_{\mathrm{PSD}},\theta_{cal},k) \\
        B=B(b, \theta_{leak,1},\theta_{leak,2}, \theta_{add},p_{PSD,bkg},\theta_{cal})
    \end{aligned}
\end{equation}

Here, $s$ is the signal strength to scale the predicted CE$\nu$NS spectrum. The upper limit is determined by scanning over a wide range of values for $s$. The remaining nuisance parameters are assigned as follows: $\phi$ is the neutrino flux from the reactor, $m_{\mathrm{act}}$ is the active mass, $\sigma_{\mathrm{eff}}$ and $\mu_{\mathrm{eff}}$ are the two parameters describing the trigger efficiency, $\sigma_{\mathrm{PSD}}$ and $\mu_{\mathrm{PSD}}$ describes the signal acceptance of the pulse shape discrimination, $\theta_{cal}$ is the uncertainty of the energy scale calibration, $k$ is the quenching parameter from the Lindhard theory, $b$ scales the Monte Carlo background spectrum, $\theta_{leak,1}$  and $\theta_{leak,2}$ describe the new background component introduced in the \textsc{Conus} data after the leakage test, $\theta_{add}$ describes the additional reactor on background found in the data and $p_{PSD,bkg}$ scales the background rejection efficiency of the pulse shape discrimination cut. \\
\newline
 Details on the application of the pulse shape discrimination in the \textsc{Conus} experiment can be found in ~\cite{bonet2023pulse}. The function that describes the signal acceptance of the PSD cut in the likelihood fit is:

\begin{equation}\label{eq:PSD_sig}
    \varepsilon_{PSD,sig} = 0.97\cdot\text{erf}\left(\frac{E_{ee}/eV_{ee}-p_1}{p_2}\right)\, ,
\end{equation}

indicating a chosen CEvNS signal acceptance of 97\% for energies above 0.6\,keV$_{ee}$. The values of the parameters $p_1$ and $p_2$ can be found in table ~\ref{tab:psd}. 

\begin{table}
\caption{\label{tab:psd} Value from the fit of the PSD signal acceptance in Run-5 data for the $p_1$ and $p_2$ parameters.}
\centering
\begin{tabular}{|l|c|c|}\hline
Detector & p1 & p2 \\\hline
C1 & -0.35~$\pm$~0.19 & 0.35~$\pm$~0.10\\
C2 & -0.51~$\pm$~0.23 & 0.44~$\pm$~0.12\\
C4 & -0.43~$\pm$~0-14 & 0.43~$\pm$~0.09\\\hline
\end{tabular}
\end{table} 

The resulting background rejection of the pulse shape cut was considered by implementing the slow pulse rejection efficiency of the three detectors in the Monte Carlo simulations of the background model (see ~\cite{bonet2023pulse}). The systematic uncertainty on this efficiency is taken into account through an additional fit parameter $p_{\mathrm{PSD,bkg}}$ for each detector that scales the overall background rejection.

The trigger efficiency was determined by injecting artificial signals produced by a pulse generator with the same rise time as of the physical signals. A detailed scan allowed to measure the detector response as a function of the energy. The description of the experimental trigger efficiency curves was obtained using the fit function:
\begin{equation}\label{eq:trigger_efficiency}
    \varepsilon_{trig} = 0.5\cdot\left(1+\text{erf}\left(\frac{E_{ee}/eV_{ee}-t_1}{t_2}\right)\right)\, .
\end{equation}

The development of the trigger efficiency curve parameters during Run-5 is shown in figure~\ref{fig:4_2}, with differences of less than 2\% over the whole run. The values obtained from the fit of all measurements during Run-5 combined are summarized in table~\ref{tab:trigger}.  

\begin{table}
\caption{\label{tab:trigger} Value from the fit of the combined trigger efficiency measurements during Run-5 for the $t_1$ and $t_2$ parameters.}
\centering
\begin{tabular}{|l|c|c|}\hline
Detector & $t_1$ & $t_2$ \\\hline
C1 & 223~$\pm$~7 & 104~$\pm$~9\\
C2 & 274~$\pm$~6 & 128~$\pm$~9\\
C4 & 283~$\pm$~3 & 132~$\pm$~5\\\hline
\end{tabular}
\end{table} 

\begin{figure*}
    \includegraphics[width=0.45\textwidth]{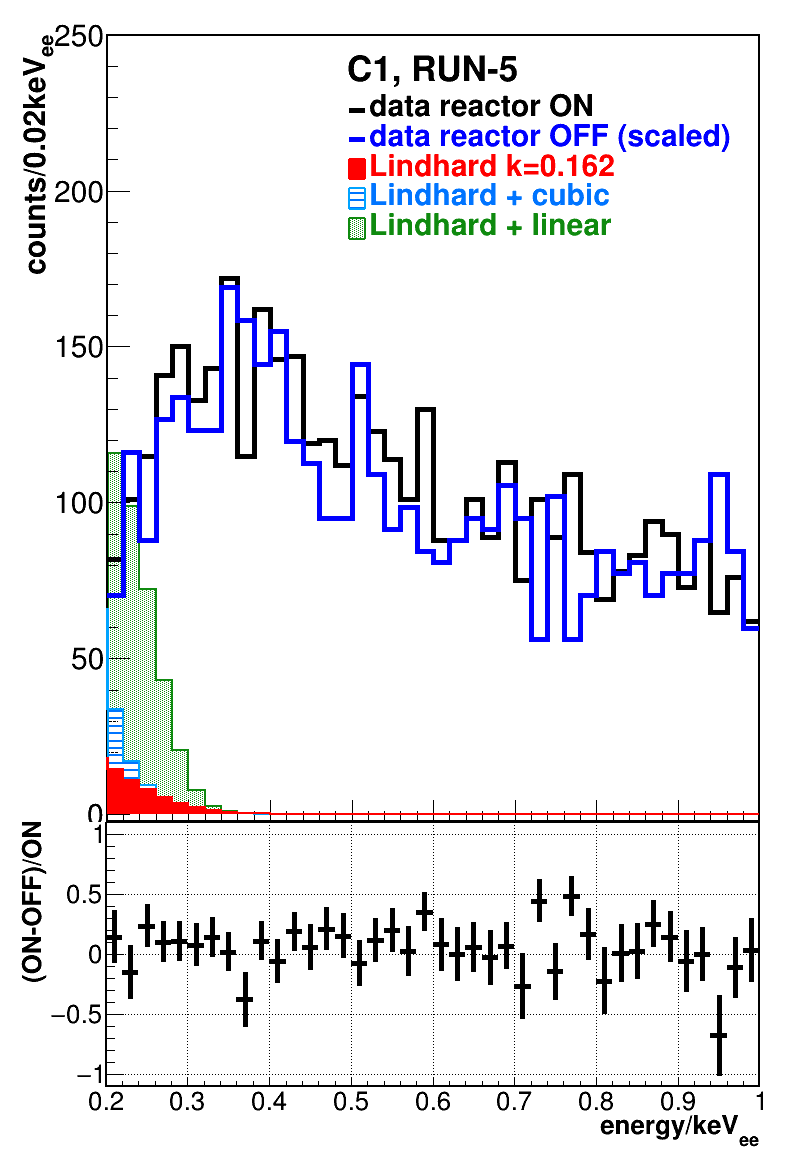}
    \hspace{1.0cm}
	\includegraphics[width=0.45\textwidth]{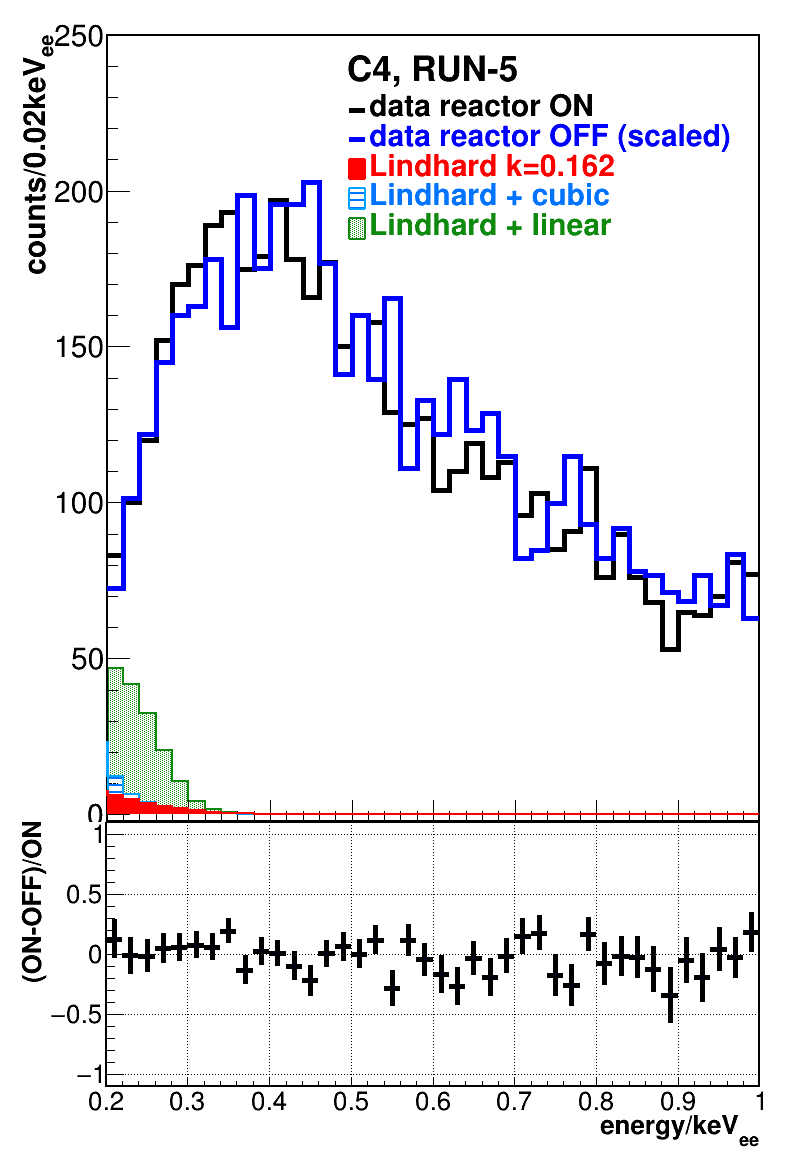}
	\caption{Reactor on vs reactor off data for the C1 (left) and C4 (right) detectors taken in Run-5 of the \textsc{Conus} experiment. In addition, the predicted CE$\nu$NS events are shown for the different quenching descriptions.}
	\label{fig:8}
\end{figure*}

Lastly, figure~\ref{fig:8} shows the reactor on and reactor off spectra acquired with the C1 and C4 respectively, including the upper limit on the CE$\nu$NS signal as obtained from the described likelihood fit.

\subsection*{Stability of environmental data}

\begin{figure*}
	\includegraphics[width=0.4\textwidth]{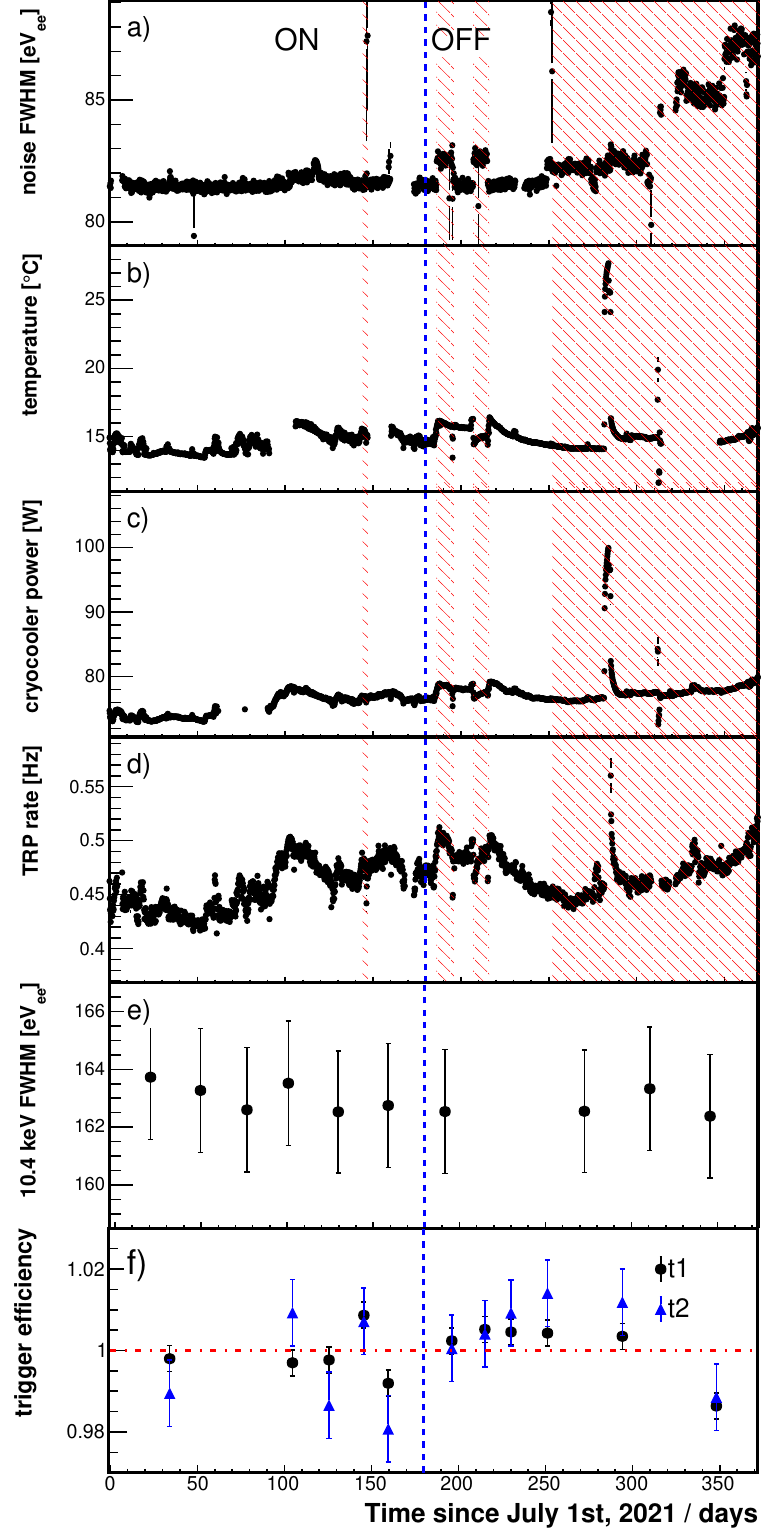}
 \hspace{1.0cm}
 	\includegraphics[width=0.4\textwidth]{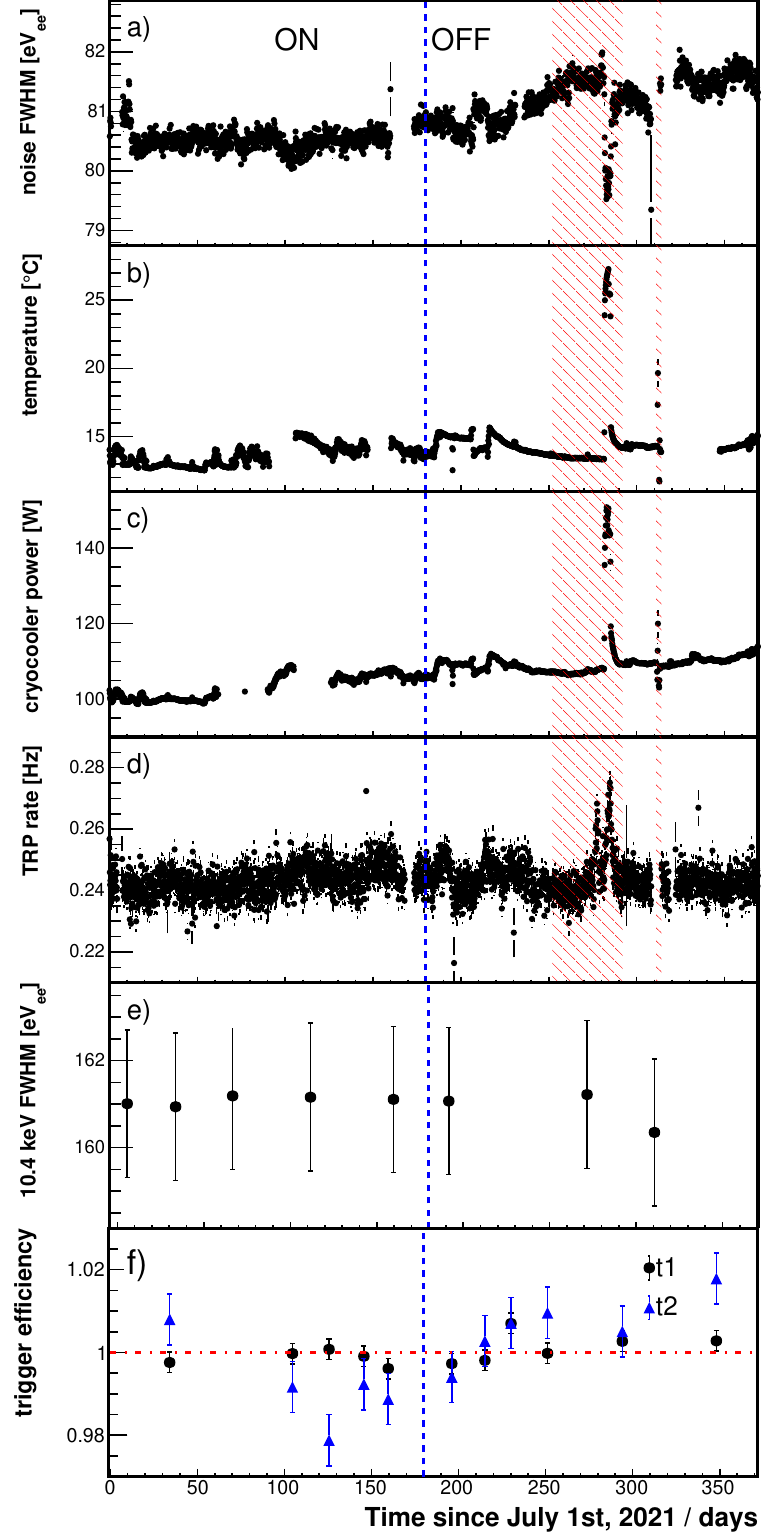}
	\caption{Evolution of parameters for the C1 (left) and C4 (right) detectors  during Run-5. From top to bottom: a) FWHM of noise peak, b) air temperature close to the detector preamplifier, c) cryocooler power consumption, d) TRP trigger rate, e) FWHM of K-shell EC peak of $^{71}$Ge, f) relative variation of trigger efficiency curve parameters~\cite{bonet2023pulse}.}
	\label{fig:4_2}
\end{figure*}

The evolution of the main environmental parameters during reactor on and off is shown in figure~\ref{fig:4_2}-left and figure~\ref{fig:4_2}-right, for the C1 and C4 detectors respectively. The excluded data set during reactor off is significantly longer for the C1 detector as compared to C2, due to the unstable noise conditions for this detector. The shape of the noise peak arising close to the lower end of the ROI was found to follow a Gaussian distribution and its width (FWHM) was monitored over time. Time periods with variations of the FWHM over 1\,eV$_{ee}$ were removed (red-shadowed area). The stability of the peak width (FWHM) of the peak at 10.37\,keV$_{ee}$ from K-shell electron capture in $^{71}$Ge is also displayed with variations below 2\,eV$_{ee}$ for both detectors. 

Overall, all the measures undertaken to improve the stability of the environment parameters for the final data taking run at KBR enabled us to collect more exposure than during any of the previous runs.  

\subsection*{Summary}
In this appendix, we provided details about the background model of the CONUS Run-5 dataset. It was obtained using a new DAQ system with highly reduced electronic noise contributions and more sophisticated trigger algorithms as in previous datasets. We could demonstrate that our physical background model explains our reactor off data and that there is no room for an additional significant component from noise events or unknown origin. We explicitly described the fit functions of the quenching models used in this Run-5 analysis and made them available to the community. The results of the individual likelihood fits for the single detectors and the associated spectra are fully consistent. An emphasis during the Run-5 data taking was control and stability of the environmental parameters, which allowed to reduce microphonic noise originating mainly from the detector cryocoolers.
\end{document}